\documentstyle[aps,prd,eqsecnum,epsf]{revtex}

\bibstyle{unsrt}
\begin{document}

\preprint{LA-UR-99-3026}

\title{Variational Ansatz for ${\cal PT}$-Symmetric Quantum Mechanics}

\author{Carl M. Bender$^1$, Fred Cooper$^{2,3}$, Peter N. Meisinger$^1$,
and Van M. Savage$^1$}
\address{${}^1$Department of Physics, Washington University, St. Louis,
MO
63130, USA}
\address{${}^2$Department of Physics, Boston College, Chestnut Hill, MA
02167, USA}
\address{${}^3$T8 MS B285, Theoretical Division, Los Alamos National
Laboratory, Los Alamos, NM 87545, USA}
\date{\today}

\maketitle
 
\begin{abstract}
A variational calculation of the energy levels of the class
of ${\cal PT}$-invariant quantum mechanical models described by the
non-Hermitian Hamiltonian $H=p^2-(ix)^N$ with $N$ positive and $x$
complex
is presented. The energy levels are determined by finding the stationary
points
of the
functional $\langle H\rangle(a,b,c)\equiv\left(\int_C
dx\,\psi(x)H\psi(x)\right)
/\left(\int_C dx\,\psi^2(x)\right)$, where 
$\psi(x)=(ix)^c\exp\left(a(ix)^b\right)$
is a three-parameter class of ${\cal PT}$-invariant
trial wave functions. The integration contour $C$ used to define
$\langle H\rangle(a,b,c)$ lies
inside a wedge in the complex-$x$ plane in which the wave function falls
off
exponentially at infinity. Rather than having a local minimum the
functional has
a saddle point in the three-parameter $(a,b,c)$-space. At this saddle
point the
numerical prediction for the ground-state energy is extremely accurate
for a
wide range of $N$. The methods of supersymmetric quantum mechanics are
used
to determine approximate wave functions and energy eigenvalues of the
excited
states of this class of non-Hermitian Hamiltonians.
\end{abstract}

 
\section{Introduction}
\label{s1}

In a recent letter \cite{r1} the spectra of the class of
non-Hermitian ${\cal PT}$-symmetric Hamiltonians of the form 
\begin{equation}
H=p^2-(ix)^N\quad(N\geq2)
\label{eq1.1}
\end{equation}
were shown to be real and positive. It has been conjectured that the
reality and
positivity of the spectra are a consequence of ${\cal PT}$ symmetry.

The Schr\"odinger differential equation corresponding to the eigenvalue
problem $H\psi=E\psi$ is 
\begin{equation}
-\psi''(x)-(ix)^N\psi(x)=E\psi(x).
\label{eq1.2}
\end{equation}
To obtain real eigenvalues from this equation it is necessary to define
the
boundary conditions properly. This was done in Ref.~\cite{r1} by
analytically
continuing in the parameter $N$ away from the harmonic oscillator value
$N=2$.
This analytic continuation defines the boundary conditions in the
complex-$x$
plane. The regions in the cut complex-$x$ plane in which $\psi(x)$
vanishes
exponentially as $|x|\to\infty$ are {\it wedges}. The wedges
for $N>2$ were chosen to be the continuations of the wedges for the
harmonic oscillator, which are centered about the negative and positive
real
axes and have angular opening ${\pi\over2}$. For arbitrary $N>2$ the
anti-Stokes'
lines at the centers of the left and right wedges lie below the real
axis
at the angles
\begin{eqnarray}
\theta_{left}&=&-\pi+\left({N-2\over N+2}\right){\pi\over2},\nonumber\\
\theta_{right}&=&-\left({N-2\over N+2}\right){\pi\over 2}.
\label{eq1.3}
\end{eqnarray}
The opening angle of these wedges is ${2\pi\over N+2}$. In
Ref.~\cite{r1} the
time-independent Schr\"odinger equation was integrated numerically
inside the
wedge to determine the eigenvalues to high precision.
Observe that as $N$ increases from its harmonic oscillator value
($N=2$), the
wedges bounding the
integration path undergo a continuous deformation as a function of $N$.

In this work we show that the ground-state energy for the ${\cal
PT}$-invariant
quantum mechanical models in Eq.~(\ref{eq1.1}) can be obtained from the
condition that the functional
\begin{equation}
E(a,b,c,\ldots)=\langle H\rangle(a,b,c)\equiv{\int_C
dx\,\psi(x)H\psi(x)\over
\int_C dx\,\psi^2(x)},
\label{eq1.4}
\end{equation}
be stationary as a function of the variational parameters
$a,~b,~c,~\ldots$
of the trial ground-state wave function $\psi(x)$. In
Eq.~(\ref{eq1.4}) the integration
contour $C$ lies in appropriate wedges in the complex plane, as
explained in
Sec.~\ref{s2}.

We obtain variational approximations to the higher eigenvalues and wave
functions using the following procedure \cite{r2,r3}: First, we obtain a
${\cal PT}$-symmetric superpotential from the approximate
ground-state trial wave function. From this we construct
a supersymmetric ${\cal PT}$-symmetric
partner potential. Next, we use variational methods to find the
ground-state energy and wave function of the Hamiltonian associated with
the
partner potential and, from
that, a second superpotential. Iterating this process, we determine the
higher-energy eigenvalues and
the associated excited states of the original Hamiltonian.

We now review this procedure in more detail.
Let us consider the Hamiltonian $H^{(0)}$ whose eigenvalues $E_n^{(0)}$
and eigenfunctions $\psi_n^{(0)}$ are labeled by the index $n$.
Subtracting the
ground-state energy $E_0^{(0)}$ from the Hamiltonian $H^{(0)}$ gives a
new
Hamiltonian $H^{(1)}\equiv H^{(0)}-E_0^{(0)}$. The eigenvalues of this
new
Hamiltonian are shifted accordingly, $E_n^{(1)}=E_n^{(0)}-E_0^{(0)}$,
but the
eigenfunctions of $H^{(1)}$ remain unchanged:
$\psi_n^{(1)}(x)=\psi_n^{(0)}(x)$.

By construction, $H^{(1)}$ has zero ground-state energy. Thus, we may
regard
$H^{(1)}$ as the first component of a two-component supersymmetric
${\cal PT}$-symmetric Hamiltonian that can be written in factored form
\cite{r4}:
\begin{eqnarray}
H^{(1)}&=&A_-^{(1)}A_+^{(1)}\nonumber\\
&=&[{1\over i}{d\over dx}-W^{(1)}(x)][{1\over i}{d\over dx}+W^{(1)}(x)].
\label{eq1.5}
\end{eqnarray}
The ground-state wave function of $H^{(1)}$ satisfies the first-order
differential equation
\begin{equation}
[{1\over i}{d\over dx}+W^{(1)}(x)]\Psi_0^{(1)}(x)=0.
\label{eq1.6}
\end{equation}
{}From this equation we determine the superpotential $W^{(1)}$:
\begin{equation}
W^{(1)}(x)=-{{\Psi_0^{(1)}}'(x)\over i\Psi_0^{(1)}(x)}=
-{{\Psi_0^{(0)}}'(x)\over i\Psi_0^{(0)}(x)}.
\label{eq1.7}
\end{equation}

The supersymmetric partner of $H^{(1)}$, which we call $H^{(2)}$, is
given by
the product of the operators $A_\pm^{(1)}$ in the opposite order from
that in
Eq.~(\ref{eq1.5}):
\begin{eqnarray}
H^{(2)}&=&A_+^{(1)}A_-^{(1)}\nonumber\\
&=&-{d^2\over dx^2}-\left[W^{(1)}(x)\right]^2-{1\over i}\left[W^{(1)}(x)
\right]'.
\label{eq1.8}
\end{eqnarray}

Note that the spectrum of $H^{(2)}$ is the same as the spectrum of
$H^{(1)}$
except that it lacks a state corresponding to the ground state of
$H^{(1)}$.
It is easy to verify that the eigenstates of $H^{(2)}$ are explicitly
given by
$\psi_n^{(2)}=A_+^{(1)}\psi_{n+1}^{(1)}$ and the corresponding
eigenvalues are
$E_n^{(2)}=E_{n+1}^{(1)}$. Thus, if we can find the ground-state wave
function and energy of $H^{(2)}$, then we can use these quantities to
express
the first excited wave function and energy of $H^{(0)}$:
\begin{eqnarray}
\psi_1^{(0)}(x)&=&{A_-^{(1)}\psi_0^{(2)}(x)\over E_0^{(2)}},\nonumber\\
E_1^{(0)}&=&E_0^{(2)}+E_0^{(0)}.
\label{eq1.9}
\end{eqnarray}

By iterating this procedure we can construct a hierarchy of
Hamiltonians $H^{(m)}$ ($m=1,~2,~3,~\ldots$). The successive energy
levels of
$H^{(0)}$ are then obtained by finding the ground-state wave functions
and
energies of $H^{(2m)}$ using variational methods that we develop in
Sec.~\ref{s2}. It is important to realize that because the higher wave
functions
of $H^{(0)}$ require an increasing number of derivatives, we need
increasingly
accurate ground-state wave functions to obtain a reasonable
approximation to the
higher eigenvalues. A more extensive discussion of this procedure can be
found
in Refs.~\cite{r2,r3}.

This paper is organized very simply. In Sec.~\ref{s2} we use variational
methods
to calculate the ground-state energy and wave function of the
Hamiltonian $H$ in
Eq.~(\ref{eq1.1}) for various values of $N$. Then in Sec.~\ref{s3} we
calculate the first few excited energy levels and wave functions. In
Sec.~\ref{s4} we make some concluding remarks and suggest some
areas for future investigation.

\section{Variational Ansatz for ${\cal PT}$-Symmetric Hamiltonians}
\label{s2}

In previous variational calculations \cite{r3} two-parameter
``post-Gaussian''
nodeless trial wave functions of the general form
\begin{equation}
\psi(x)=\exp\left(-a|x|^b\right)
\label{eq2.1}
\end{equation}
were used to obtain estimates for the ground states of the hierarchy of
Hamiltonians constructed from the anharmonic oscillator $p^2+x^4$. In
this
paper we obtain estimates for
the ground states of the hierarchy of ${\cal PT}$-symmetric Hamiltonians
arising from $H$ in Eq.~(\ref{eq1.1}).

The concept of ${\cal PT}$ symmetry is discussed in detail in
Refs.~\cite{r1}
and \cite{r5}. To be precise, a function $F(x)$ of a complex argument
$x$ is
${\cal PT}$ symmetric if
\begin{equation}
[F(x)]^*=F(-x^*).
\label{eq2.3}
\end{equation}
Thus, any real function of $ix$ is manifestly ${\cal PT}$ symmetric. In
previous numerical calculations \cite{r1} it was found that the
eigenfunctions
of the ${\cal PT}$-symmetric Hamiltonians in Eq.~(\ref{eq1.1}) were {\it
all}
${\cal PT}$ symmetric. Thus, for our variational calculation we choose a
three-parameter trial wave function that is explicitly ${\cal PT}$
symmetric:
\begin{equation}
\psi(x)=(ix)^c\exp\left(a(ix)^b\right),
\label{eq2.2}
\end{equation}
where the variational parameters $a$, $b$, and $c$ are real.

The most notable difference between the Hermitian trial wave function in
Eq.~(\ref{eq2.1}) and the ${\cal PT}$-symmetric trial wave function in
Eq.~(\ref{eq2.2}) is the number of variational parameters. It is not
advantageous to have a prefactor in Eq.~(\ref{eq2.1}) of the form
$|x|^c$
because the expectation value of the kinetic term $-d^2/dx^2$ of the
Hamiltonian
does not exist unless $c>{1\over2}$. Because of the singularity at $x=0$
we must
exclude the case $c\leq{1\over2}$ in conventional variational
calculations.
However, positive values of $c$ are in conflict with the large-$|x|$
behavior of
the WKB approximation to the exact wave function; WKB theory predicts a
negative
value for the parameter $c$.

We emphasize strongly that in conventional variational calculations the
integration path is the real axis. For the case of $\psi(x)$ in
Eq.~(\ref{eq2.1}) one cannot deform the contour of integration to avoid
the
origin because the function $|x|$ is not analytic. More generally, the
notion
of path independence is not applicable in conventional variational
calculus
because the functional to be minimized involves an integral over the
{\it
absolute square} $\psi^*(x)\psi(x)$ of the trial wave function. For the
case
of ${\cal PT}$-symmetric quantum mechanics the trial wave function
$\psi(x)$ in
Eq.~(\ref{eq2.2}) is analytic in the cut $x$ plane. (The cut is taken to
run
along the positive-imaginary axis; this choice is required by ${\cal
PT}$
symmetry.) Moreover, the functional involves an integral over the
square,
not the absolute square, of the wave function \cite{r5}. These two
properties imply that the integration
contour can be deformed so long as the endpoints of the contour lie
inside of
the wedge in Eq.~(\ref{eq1.3}). In particular, 
the contour may be chosen to avoid the infinite singularity at the
origin
that occurs when the parameter $c$ in Eq.~(\ref{eq2.2}) is negative.
Thus, in ${\cal PT}$-symmetric quantum mechanics we
can include the additional parameter $c$ anticipating that we will
obtain highly accurate numerical results for the energy levels.

In general, for Hermitian Hamiltonians it is well known that the
variational
method always gives an upper bound for the ground-state energy. That is,
the
stationary point of the expectation value of the Hamiltonian in the
normalized
trial wave function [see Eq.~(\ref{eq1.4})] is an {\it absolute
minimum.} By
contrast, for ${\cal PT}$-invariant quantum mechanics, too little is
known about
the state space to prove any theorems about the nature of stationary
points.
Thus, we will simply look for all stationary points of the function
$E(a,b,c)$
in Eq.~(\ref{eq1.4}), and not just minima.

The first step in our calculation is to evaluate the functional
$E(a,b,c)$
in Eq.~(\ref{eq1.4}). To do this we must make a definite choice of
contour $C$.
It is convenient to choose the
path of integration in the complex-$x$ plane to follow the rays for 
which the trial wave function in Eq.~(\ref{eq2.2}) does not oscillate
and
falls off exponentially as $\exp\left(-ar^b\right)$.
(We assume implicitly here that $a$ is positive.)
These rays are symmetrically placed about the negative-imaginary axis.
Specifically, we set
\begin{equation}
z=re^{i\theta_L},~~{\rm where}~~\theta_L=-{\pi\over 2}-{\pi\over b}
\label{eq2.4}
\end{equation}
on the left side of the imaginary axis, with the path running from
$r=-\infty$
to $r=0$ ($\theta_L$ fixed), and
\begin{equation}
z=re^{i\theta_R},~~{\rm where}~~\theta_R=-{\pi\over2}+{\pi\over b}
\label{eq2.5}
\end{equation}
on the right side of the imaginary axis, with the path running from
$r=0$ to
$r=\infty$ ($\theta_R$ fixed). Note that when
$b=2$, the path lies on the real axis; for $b\neq2$, the path develops
an
elbow at the origin.

Our key assumption is that this integration path lie inside the
asymptotic
wedges described in Eq.~(\ref{eq1.3}). This requirement restricts the
value of
$b$:
\begin{equation}
{1\over3}(N+2)\leq b\leq N+2.
\label{eq2.6}
\end{equation}

Now, we evaluate the integral in Eq.~(\ref{eq1.4}) in terms of Gamma
functions:
\begin{eqnarray}
E(\alpha,\beta,\gamma)&=&{(\beta-\beta^2-\gamma)\Gamma(1-\beta-\gamma)\alpha^{2
\beta}\over4\beta^2\Gamma(1+\beta-\gamma)}\nonumber\\
&&\qquad-{\Gamma(1-\beta-\gamma)\alpha^{-N\beta}\over
\Gamma(1-\beta-N\beta-\gamma)},
\label{eq2.7}
\end{eqnarray}
where
\begin{eqnarray}
\alpha=2a,\quad\beta=1/b,\quad\gamma=2c/b.
\label{eq2.8}
\end{eqnarray}

\begin{figure*}[h]
\vspace{6.0in}
\includegraphics{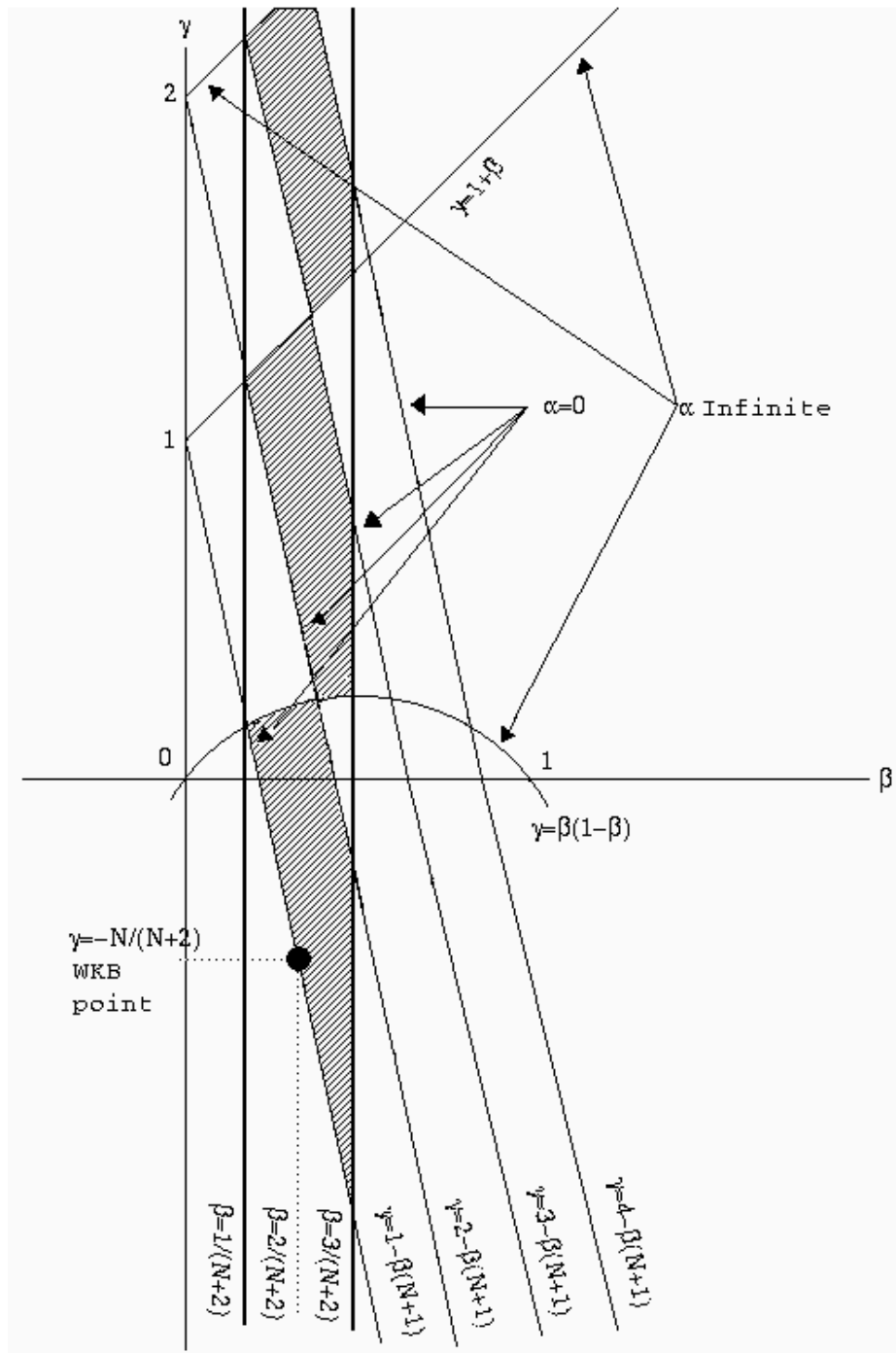}
\caption{Regions (shown as shaded) of acceptable values of $\beta$ and
$\gamma$. These values of $\beta$ and $\gamma$ are acceptable because they give
positive $\alpha$  in Eq.~(\ref{eq2.9}).}
\label{f1}
\end{figure*}

To find a stationary point of the function $E(\alpha,\beta,\gamma)$ we
must
calculate its partial derivatives with respect to the parameters
$\alpha$,
$\beta$, and $\gamma$, and determine the values of these parameters for
which
the partial derivatives vanish. It is simplest to calculate the
derivative with
respect to $\alpha$; requiring that this derivative be zero gives an
expression
for $\alpha$ in terms of $\beta$ and $\gamma$:
\begin{equation}
\alpha^{(2+N)\beta}={2N\beta^2\Gamma(1+\beta-\gamma)\over
(\beta^2-\beta+\gamma)\Gamma(1-\beta-N\beta-\gamma)}.
\label{eq2.9}
\end{equation}
Recalling that the parameter $a$, and therefore $\alpha$, must be
positive gives
regions in the $(\beta,\gamma)$ plane of the allowable values of $\beta$
and
$\gamma$. These regions are shown as shaded areas in Fig.~\ref{f1}.

There are many distinct regions for which $\alpha$ is
positive. We describe the boundaries of
these regions in detail. First, all of these regions must
satisfy the inequality
\begin{equation}
{1\over N+2}\leq\beta\leq{3\over N+2},
\label{eq2.10}
\end{equation}
which follows from Eq.~(\ref{eq2.6}). These bounds are shown as two
heavy
vertical lines in Fig.~\ref{f1}.

Second, there is an inverted parabola
\begin{equation}
\gamma=\beta-\beta^2.
\label{eq2.11}
\end{equation}
The lowest region is bounded above by this parabola and the next higher
region 
is bounded below by this parabola.

Third, there is an infinite sequence of downward-sloping straight lines
of the
general form
\begin{equation}
\gamma=k-\beta(N+1)\qquad(k=1,~2,~3,~\ldots).
\label{eq2.12}
\end{equation}
These lines bound the shaded regions on the left and on the right.

Fourth, there is an infinite sequence of upward-sloping straight lines
of the
general form
\begin{equation}
\gamma=k+\beta\qquad(k=1,~2,~3,~\ldots).
\label{eq2.13}
\end{equation}
These lines are upper and lower bounds of the shaded regions.

WKB theory predicts that
\begin{eqnarray}
\alpha={4\over N+2},\quad\beta={2\over N+2},\quad\gamma=-{N\over N+2}.
\label{eq2.14}
\end{eqnarray}
The special WKB values $(\beta={2\over N+2},\gamma=-{N\over N+2})$ are
indicated
by a heavy dot in Fig.~\ref{f1}. Note that this point lies on the
boundary of
the only shaded region that lies below the $\beta$ axis.
We will restrict our attention to the stationary point (there is only
one)
of $E(\alpha,\beta,\gamma)$ that lies in this special region because it
is
nearest to
the WKB point. We will see that the stationary point in this special
region is a
local maximum as a function of $\beta$ and $\gamma$,
and not a local minimum as in the case of conventional Hermitian
theories.

To locate the stationary point in the $(\beta,\gamma)$ plane
we solve Eq.~(\ref{eq2.9})
for $\alpha$ and substitute the result back into the energy functional
$E(\alpha,\beta,\gamma)$ in Eq.~(\ref{eq2.7}). We then construct a
highly-accurate contour plot of the resulting function of $\beta$ and
$\gamma$.
{}From our analysis we are able to locate the stationary points to four
significant digits for various values of $N$. Our results for $N=3$,
$N=4$,
and $N=5$ are shown in Tables \ref{t1} and \ref{t2}. In Table \ref{t3}
we show
that in the $(\beta,\gamma)$ plane, the stationary point is a local
maximum.

To verify the accuracy of the ground-state trial wave function we have
calculated the normalized expectation value of $x^P$ using $\psi(x)$ in
Eq.~(\ref{eq2.2}):
\begin{eqnarray}
{\langle0|~x^P~|0\rangle\over\langle0|0\rangle}&=&{\int_C
dx\,x^P\psi^2(x)\over
\int_C dx\,\psi^2(x)}\nonumber\\
&=&
(-i)^P{\Gamma(1-\beta-\gamma)\alpha^{-P\beta}\over\Gamma(1-\beta-P\beta
-\gamma)}.
\label{eq2.15}
\end{eqnarray}
Substituting the values of $\alpha$, $\beta$, and $\gamma$ taken from
Table
\ref{t1} for $N=3$, $N=4$, and $N=5$ we display in Table \ref{t4} the
expectation values of $x^P$ for $P=1,~2,~\ldots,~5$. To measure the
accuracy of
our variational wave function we can check the numbers in Table \ref{t4}
in two
ways. First, we have calculated the exact values of
$\langle0|~x~|0\rangle/
\langle0|0\rangle$ for $N=3$, $N=4$, and $N=5$. The values are
$-0.590073i$,
$-0.866858i$, and $-1.013102i$, respectively. These numbers differ
from those in the first row in Table \ref{t4} by about one part in a
thousand.
Second, combining the operator equation of motion obtained from $H$ in 
Eq.~(\ref{eq1.1}),
\begin{equation}
x''(t)=2iN(ix)^{N-1},
\label{eq2.16}
\end{equation}
with the time-translation invariance of $\langle0|~x~|0\rangle$, 
we find that the expectation value of $x^{N-1}$ vanishes for an $(ix)^N$
theory. This is evident to a high degree of accuracy in Table \ref{t4}.

One further check of the accuracy of the trial wave function is to
compare
its asymptotic behavior with that of the WKB approximation to $\psi(x)$.
For
example, from Eq.~(\ref{eq2.14}) the WKB value of $\beta$ for $N=3$ is
$0.4$.
For a simpler two-parameter trial wave function in which we set
$\gamma=0$,
the stationary point occurs at $\beta=0.36$. For the three-parameter
case we
get $\beta=0.3855$, which is much closer to the value $0.4$. Indeed, all
of the
values in Table \ref{t1} are very close to the WKB values in
Eq.~(\ref{eq2.14}).

\section{Higher Energy Levels}
\label{s3}

We follow the procedure outlined in Sec.~\ref{s1} and use trial wave
functions of the form in Eq.~(\ref{eq2.2}) to construct the
first few excited states. As one might have
expected, the numerical accuracy decreases for the higher states. The
ground-state values of $(\alpha,\beta,\gamma)$ for the case $N=3$ are
$(0.72066,0.3855,-0.157)$, as shown in Table \ref{t1}. For the first
excited
state we obtain the values
$(0.55750,0.3568,-0.329)$ and for the second excited state we obtain
$(0.39925,0.3255,-0.466)$. Using these parameters we predict that the
first
excited energy level $E_1$ is $4.11738$ and that the second excited
energy level
$E_2$ is $7.53886$. The exact values of $E_1$ and $E_2$ obtained by
numerical
integration of the Schr\"odinger equation (\ref{eq1.2}) (see
Ref.~\cite{r1})
are $4.1092$ and $7.5621$. Thus, the relative errors in our calculation
of
$E_0$, $E_1$, and $E_2$ are roughly 1 part in 2000, 4 parts in 2000, and
7 parts
in 2000.

For the case $N=4$ the ground-state values of $(\alpha,\beta,\gamma)$
are
$(0.5800,0.3185,-0.2745)$ (see Table \ref{t1}). For the first excited
state we
obtain $(0.41763,0.2915,-0.614)$. From these parameters we predict that
the
first excited energy level $E_1$ is $6.03769$. The exact value of $E_1$
is
$6.0033$. Thus, the relative error in this calculation of $E_0$ and
$E_1$ is
roughly 3 parts in 5000 and 25 parts in 5000.

\begin{table}
\caption[t1]{Variational calculation of the ground-state energy for the
${\cal PT}$-symmetric Hamiltonian in Eq.~(\ref{eq1.1}) for three values
of $N$.
The stationary point of $E(\alpha,\beta,\gamma)$ for the three-parameter
trial
wave function in Eq.~(\ref{eq2.2}) is given for $N=3$, $N=4$, and
$N=5$.}
\begin{tabular}{lddd}
$N$ & $\alpha$ & $\beta$ & $\gamma$
\\ \tableline
3 & 0.7207 & 0.3855 & -0.1570 \\
4 & 0.5800 & 0.3185 & -0.2745 \\
5 & 0.4895 & 0.2720 & -0.3610 \\
\end{tabular}
\label{t1}
\end{table}

\begin{table}
\caption[t2]{Variational calculation of the ground-state energy for the
${\cal PT}$-symmetric Hamiltonian in Eq.~(\ref{eq1.1}) for three values
of $N$.
Using the stationary point of $E(\alpha,\beta,\gamma)$ taken from Table
\ref{t1} we calculate the variational value of the ground-state energy
$E_{\rm var}$ and compare it with the exact value $E_{\rm exact}$
obtained by
direct numerical integration of the Schr\"odinger equation
in (\ref{eq1.2}). Note that the relative error for $N=3$, $N=4$, and
$N=5$ is
approximately one part in $2000$.}
\begin{tabular}{lddd}
$N$ & $E_{\rm var}$ & $E_{\rm exact}$ & rel. error
\\ \tableline
3 & 1.156754 & 1.156267 & 0.042\% \\
4 & 1.478023 & 1.477149 & 0.062\% \\
5 & 1.909382 & 1.908265 & 0.059\% \\
\end{tabular}
\label{t2}
\end{table}

\begin{table}
\caption[t3]{The stationary point of
$E[\alpha(\beta,\gamma),\beta,\gamma]$ is a
local maximum in the $(\beta,\gamma)$ plane. To verify this the value of
$E$ for
$N=4$ is given at the stationary point $\beta=0.3185$, $\gamma=-0.2745$
and at
four nearby points. In a conventional Hermitian theory the stationary
point
would be a local minimum.}
\begin{tabular}{ddd}
$\beta$ & $\gamma$ & $E(\alpha,\beta,\gamma)$
\\ \tableline
0.3185 & -0.2745 & 1.478023 \\
0.3185 & -0.2750 & 1.478022 \\
0.3185 & -0.2740 & 1.478022 \\
0.3186 & -0.2745 & 1.478022 \\
0.3184 & -0.2745 & 1.478022 \\
\end{tabular}
\label{t3}
\end{table}

\begin{table}
\caption[t4]{Normalized expectation values of $x^P$ in the ground state.
These expectation values were
calculated from the trial wave function $\psi(x)$ in Eq.~(\ref{eq2.2})
with
$\alpha$, $\beta$, and $\gamma$ given in Table \ref{t1} for $N=3$,
$N=4$, and
$N=5$.}
\begin{tabular}{lddd}
$P$ & ${\langle0|~x^P~|0\rangle\over\langle0|0\rangle}\bigm|_{N=3}$ &
${\langle0|~x^P~|0\rangle\over\langle0|0\rangle}\bigm|_{N=4}$ &
${\langle0|~x^P~|0\rangle\over\langle0|0\rangle}\bigm|_{N=5}$
\\ \tableline
1 & -0.590686$i$ & -0.867264$i$ & -1.013266$i$ \\
2 & -0.000767 & -0.518220 & -0.864749 \\
3 & -0.462705$i$ & 0.000869$i$ & 0.518217$i$ \\
4 & -0.385363 & -0.492631 & 0.002080 \\
5 & -0.363227$i$ & 0.635997$i$ & 0.545538$i$ \\
\end{tabular}
\label{t4}
\end{table}

Having verified that the variational calculation of the energies of the
excited states is extremely accurate, we turn to the wave functions. We
have
performed two tests of the accuracy of the variational wave functions.
First, we have computed the integral of the square of the wave
functions.
In general, we expect that the integral of the square of the $n$th
excited wave
function to be a real number having the sign pattern $(-1)^n$. This is
precisely what we find for the $N=3$ wave functions
($\psi_0$, $\psi_1$, and $\psi_2$) and the $N=4$ wave functions
($\psi_0$ and $\psi_1$) that we have studied. To understand the origin
of this
alternating sign pattern recall the form of the wave functions for the
harmonic 
oscillator [the $N=2$ case of Eq.~(\ref{eq1.1})]. The harmonic
oscillator is
${\cal PT}$ symmetric and all of its eigenfunctions can be made
${\cal PT}$ symmetric upon multiplying by the appropriate factor of $i$.
The
first three eigenfunctions in explicit ${\cal PT}$-symmetric form are,
apart from a real multiplicative constant,
\begin{eqnarray}
\psi_0(x)&=&e^{(ix)^2/2},\nonumber\\
\psi_1(x)&=&(ix)e^{(ix)^2/2},\nonumber\\
\psi_2(x)&=&[2(ix)^2+1]e^{(ix)^2/2}.\nonumber
\end{eqnarray}
These wave functions exhibit precisely this $(-1)^n$ behavior.

Second, we have found the nodes of the variational wave functions. For
$N=3$ the
ground-state variational wave function is nodeless and the first excited
state
variational wave function has one node at $x=-0.703i$. These results are
extremely accurate: The exact ground-state wave function is also
nodeless
and the exact first-excited-state wave function has one node at
$x=-0.533i$.
We observe the same qualitative features for the case $N=4$; here,
the node of the first excited state, as determined by our variational
calculation, is located at $x=-0.972i$.

\section{Conclusions}
\label{s4}

The results of this investigation suggest several intriguing avenues of
research. The most surprising aspect of this variational calculation is
that
even though the stationary point is a saddle point in parameter space,
and not a minimum, this saddle point clearly gives correct physical
information. Clearly, it is the nonpositivity of the norm that accounts
for
the fact that the stationary point is a saddle point. The norm used in
this
paper is
\begin{equation}
\int_C dx\,\psi^2(x)
\label{eq4.1}
\end{equation}
rather than the conventional choice
\begin{equation}
\int_C dx\,\psi^*(x)\psi(x).
\label{eq4.2}
\end{equation}
However, the accuracy
of the variational approach suggests that we have found an extremely
powerful technique for obtaining approximate solutions to complex
Sturm-Liouville eigenvalue problems.

A possible explanation for the accuracy of our approach is that while
the
norm we are using is not positive definite in coordinate space, we
believe that it {\it is} positive definite in momentum space. The
positivity
for our choice of norm is easy to verify in the special case of the
harmonic oscillator. We believe that this positivity property does not
undergo
a sudden transition as we analytically continue the theory
into the complex plane by increasing $N$. We point out that the strong
advantage of our choice of norm is that $\psi^2(x)$ is an analytic
function
while $\psi^*(x)\psi(x)$ is not. This allows us to analytically continue
Sturm-Liouville problems into the complex plane. (We point out that
until
now, Sturm-Liouville problems have been limited to the real axis.)

An extremely interesting question to be considered in the future is the
completeness of the eigenfunctions of a complex ${\cal PT}$-symmetric
Sturm-Liouville problem.
It is easy to show that the eigenfunctions corresponding to different
eigenvalues are orthogonal:
\begin{equation}
\int_C dx\,\psi_m(x)\psi_n(x)=0\quad(m\neq n).
\label{eq4.3}
\end{equation}
However, it is not known whether the eigenfunctions form a complete set
and
what norm should be used in this context. It is clear that in order
to answer questions regarding completeness, a full understanding of the
distribution of zeros of the wave functions must be achieved. For a
Sturm-Liouville problem limited to the real axis completeness depends on
the
zeros becoming dense. For the case of a regular Sturm-Liouville problem
the
zeros interlace as they become dense. Our preliminary
numerical investigations suggest
that the zeros of the eigenfunctions of complex ${\cal PT}$-symmetric
Sturm-Liouville problems also become dense in the complex plane
and exhibit a two-dimensional 
generalization of the interlacing phenomenon.

One area that we have investigated in great detail is the extension
of these quantum mechanical variational calculations to quantum field
theory \cite{r6}. The natural technique to use is based on the
Schwinger-Dyson equations; successive truncation of these equations is
in
exact analogy with enlarging the space of parameters in the trial wave
functions. We find that the Schwinger-Dyson equations provide an
extremely
accurate approximation scheme for calculating the Green's functions and
masses of a ${\cal PT}$-symmetric quantum field theory.

Finally, we remark that the notion of probability in quantum mechanics
is connected with the positivity of the norm in Sturm-Liouville
problems.
We note that the conventional quantum mechanical norm is neither an
analytic
function nor invariant with respect to ${\cal PT}$ symmetry,
while the momentum-space version of the norm in Eq.~(\ref{eq4.1}) is
both. It would be a remarkable advance if one could
formulate a fully ${\cal PT}$-symmetric quantum mechanics.

We thank Dr.~S.~Boettcher for assistance in doing computer calculations.
This work was supported in part by the U.S. Department of Energy.

\newpage

\end{document}